# New pixelized Micromegas detector for the COMPASS experiment

**Damien Neyret**[a1], **Marc Anfreville**[a], **Yann Bedfer**[a], **Etienne Burtin**[a], **Nicole d'Hose**[a], **Arnaud Giganon**[a], **Bernhard Ketzer**[b], **Igor Konorov**[b], **Fabienne Kunne**[a], **Alain Magnon**[a], **Claude Marchand**[a], **Bernard Paul**[a], **Stéphane Platchkov**[a] and **Maxence Vandenbroucke**[b,a]

[a] *CEA Saclay DSM IRFU,*
  *91191 Gif sur Yvette Cedex, France*
  *E-mail*: damien.neyret@cea.fr

[b] *Technische Universität München, Physik Department, E18 group,*
  *D 85748 Garching, Germany*

ABSTRACT: New Micromegas (*Micro-mesh gaseous* detectors) are being developed in view of the future physics projects planned by the COMPASS collaboration at CERN. Several major upgrades compared to present detectors are being studied: detectors standing five times higher luminosity with hadron beams, detection of beam particles (flux up to a few hundred of kHz/mm², 10 times larger than for the present detectors) with pixelized read-out in the central part, light and integrated electronics, and improved robustness. Studies were done with the present detectors moved in the beam, and two first pixelized prototypes are being tested with muon and hadron beams in real conditions at COMPASS. We present here this new project and report on two series of tests, with old detectors moved into the beam and with pixelized prototypes operated in real data taking condition with both muon and hadron beams.

---

[1] Corresponding author.

# Contents

## 1. The COMPASS experiment and the Micromegas detectors

The COMPASS experiment [1] is dedicated to the study of the spin structure of the nucleon and the spectroscopy of hadrons. It takes advantage of the secondary muon and hadron beams delivered on the M2 beam line at the SPS (Super Proton Synchrotron) accelerator at CERN, with an energy range from 100 to above 200 GeV and with beam intensities reaching $10^8$ part/s. The COMPASS spectrometer is constituted by a fixed target and a two stage spectrometer for the detection and identification of particles at low and high momenta at high flux. We recall in section 1 the general features of the existing Micromegas detectors, and we present the new detectors in development in section 2.

### 1.1 The Micromegas detectors

Since the beginning of the COMPASS data taking in 2002, 12 Micromegas detectors are installed at the downstream edge of the fixed target to detect particles scattered at low angles. Micromegas detectors are fast gaseous detectors using a micro-mesh electrode separating a thick low-field ionization area, where the gas is ionized by the charged particles, and a high-field amplification gap where the electrons produced in the ionization gap are amplified and then read by micro-strips. As the amplification gap is very thin (in the order of 100 μm) the ions are quickly neutralized, the produced signals are then much shorter, in the order of 100 ns, than standard multi-wires chambers.

These Micromegas detectors [2] were developed by the CEA Saclay in order to fit the requirements of the COMPASS experiment to stand a flux up to 500 kHz/cm², with a spatial resolution better than 100 μm and low material budget. The amplification gap is 100 μm thick, while the ionization gap was increased from 3.2 to 5 mm in 2006 to operate the detectors at lower gain and reduce the discharge rate with hadron beams [3]. The drift and mesh electrodes are made of 5 µm copper meshes. The voltage applied on the amplification gap is around 400 to 420 V, giving a gain between 3000 to 6000. The active area of each detector covers 40x40 cm², with a blind disk of 5 cm diameter on the path of the beam. These detectors measure particle positions in one dimension, the 12 detectors being grouped by stations of 4 detectors each, covering X, Y, U and V (±45°) coordinates. The micro-strips are extended outside the active area by 30 cm to shift the electronic cards out of the spectrometer acceptance.

**1.2 Micromegas performances**

The COMPASS Micromegas detectors achieve very good performances fulfilling all the requirements. The detection efficiency is better than 98% at low particle flux, while at high flux it decreases to 96% due to the electronics occupancy, with a counting rate which can reach 150 kHz/channel (Figure 1) [4]. In nominal high flux conditions the spatial resolution is between 90 and 100 µm, except for detectors close to the polarized target and dipole magnetic fringe fields (up to 1 T), where the resolution is 10% larger due to the Lorentz angle effect.

The discharge rate has been minimized by using a light gas (Ne + 10% $C_2H_6$ + 10% $CF_4$) and also by running the detectors at moderate gain thanks to the low noise electronics using SFE16 chips [5]. Moreover, 110 pF capacitances decouple each strip from the others and from the electronics, the charges being drained by 1 MΩ resistors to the ground. In case of discharge, only one or a few strips are affected by the voltage variation, reducing the capacitance involved in the discharge and leading to a very short recovery time (~3.5 ms). With a high flux muon beam ($4.10^7$ muon/s on a 1.20 m $^6$LiD solid target), the discharge rate per detector is lower than 0.02 discharge/s. With a hadron beam the discharge rate reaches 0.04 to 0.1 discharge/s with a beam of $10^7$ hadron/s (Figure 2). This rather high discharge rate is tolerable but prevents from any future increase of the hadron beam flux. Higher discharge rates would reduce the detector efficiency and might deteriorate the hardware.

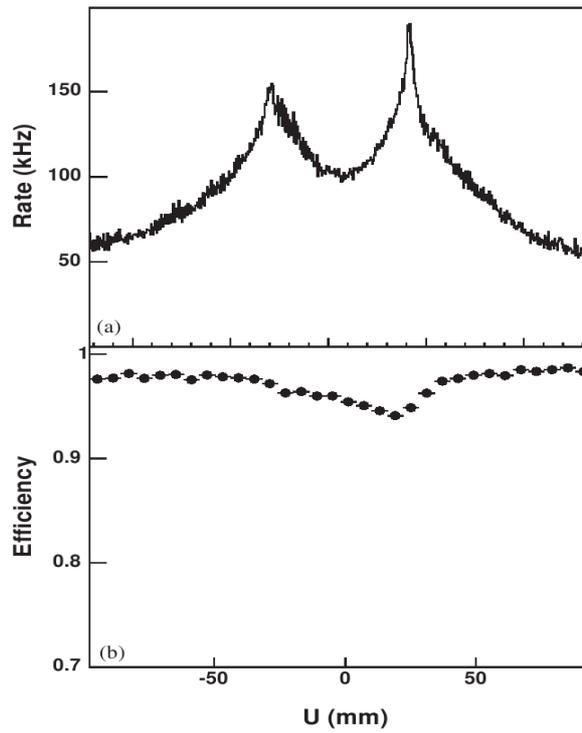

*Figure 1: Counting rate of a present Micromegas detector with high flux muon beam (upper plot) with a mesh voltage of 420 V, efficiency of the detector for these conditions (lower plot) [4]*

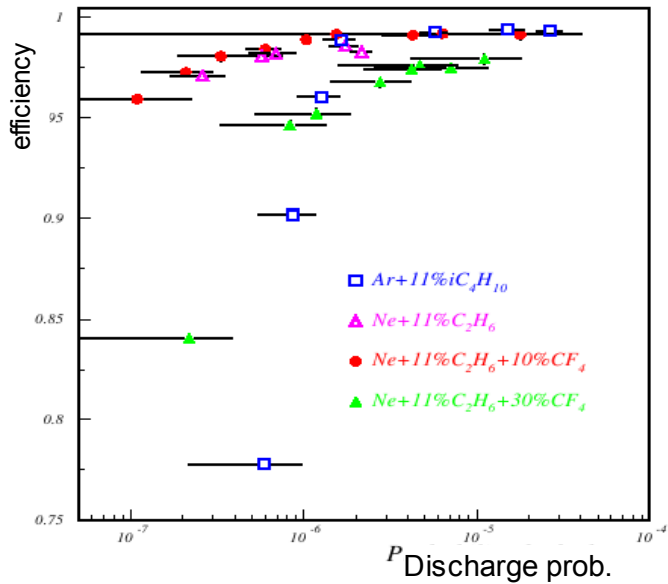

*Figure 2: Discharge probability vs efficiency for a hadron beam and for various gas mixtures [2], the highest efficiencies are reached with Ne gases.*

## 2. Project for a new pixelized Micromegas detector

The COMPASS collaboration will complete in the coming years its initial physics program [6]. For the long term future (2012 and beyond) a new proposal is in preparation, and a letter of intent was submitted to the CERN SPSC [7]. New physics items are proposed like the study of Generalized Parton Distributions via Deeply Virtual Compton Scattering, transversity via the study of the Drell-Yan reaction and further measurements on hadron spectroscopy.

The perspective to use Micromegas detectors during many more years and at a higher flux, lead us to take the opportunity to improve them. Four objectives have been defined: the new detectors must stand more than 5 times higher hadron beam flux, they must be active in the central region where the beam is crossing in order to replace present thick scintillating fiber detectors, the read-out electronics must be much lighter even with the increased number of channels, and the detectors must be more reliable and robust.

The goal of the present R&D project is to develop new Micromegas detectors which will fulfill these requirements, with 10 to 100 lower discharge rate compared to present detectors, and with a pixelized read-out for the beam area. A read-out electronics using APV25-S1 chip [8] is developed, and the new "bulk" Micromegas technology [9] is considered.

### 2.1 Discharge rate reduction

Several ideas to reduce the discharge rate are presently considered. A promising one is to cover the read-out board with a resistive coating, which would reduce the impact of the discharges by limiting the current to the strips and thus the duration of the discharges. The capacitance involved in the discharge process would also be limited by the resistive layer. Several sorts of coating are on the market: resistive foil on isolating layer, resistive coating in contact with the strips, segmented resistive coating, etc... Studies on resistive coating have already been started, in particular by sLHC and ILC CEA Saclay groups. We are joining these efforts and will study several prototypes during the next RD-51 beam tests at CERN at the end of October 2009.

The solution to add a GEM foil above the Micromegas mesh will also be studied. The charge spread and the lower Micromegas gain would lead to lower discharge rates. Such prototype will also be tested in the beam in October.

A fall-back solution would be to segment the mesh electrode. This solution would not directly decrease the discharge rate, but it would lower the capacitance involved in the discharges and thus shorten the recovery time.

### 2.2 Pixel read-out in the center of the detector

The particle flux in the center of the COMPASS Micromegas detectors is expected to be very high, of the order of 200 kHz/mm², since the incoming beam goes through the detectors here. A read-out with strips would lead to hit rates of the order of 500 kHz/channel, generating an inefficiency due to the electronics occupancy higher than 10%.

To decrease the hit rate to about 200 kHz/channel, a possible solution is to use a pixelized read-out with a pixel area of about 1 mm² in the central area. To achieve a better spatial resolution these pixels would be rectangular and parallel to the strips, with a size of 2x0.5 mm² in the center and 4x0.5 mm² at larger angles (see Figure 3). The usual strip read-out would be used outside the beam area. The number of channels necessary to read these pixels would be of the order of 1300, in addition to the 1024 strips.

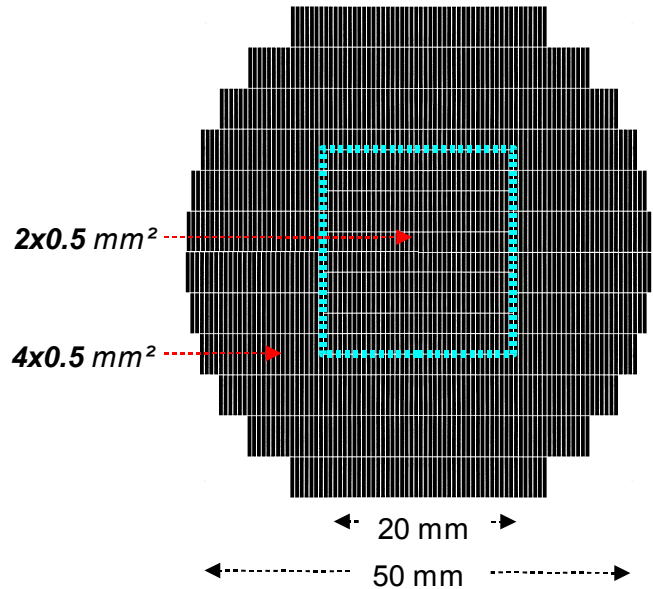

*Figure 3: Pixelized part of the future Micromegas detectors, with 2x0.5 mm² rectangular pixels in the center and 4x0.5 mm² in the external part*

**2.3 Read-out electronics with APV chips**

The present read-out electronics using SFE16 amplifiers/discriminators in association with F1 TDC chips is working well since the beginning of the COMPASS experiment. However the space taken by the electronic cards is quite important as each card reads only a small number of channels (16 for a SFE16 card, 64 for a F1 board). The power consumption is also important, around 500 W per detector, requiring a high cooling power.

As the new detectors will have twice more channels, it is not desirable to use the same electronics, due to the lack of space and the high heat production. A more compact and integrated electronics is being studied, using APV chips. The APV25-S1 is a 128-channels amplifier and analog multiplexer on a 7x8 mm² silicon chip directly bonded to the printed circuit board. The characteristics of the amplifier are tunable using configuration registers, and its timing constants can be adapted to the characteristics of the detector. The analog values are sampled at 40 MHz and when an event is triggered, a configurable number of samples (usually three samples) from each channel are multiplexed and sent to a digitization card. This card features a 10-bits flash ADC for each connected APV, and a FPGA which applies several algorithms: pedestal subtractions, common mode noise correction, and zero suppression.

APV electronic read-out is already used in the COMPASS experiment. It was developed by the E18 group of the Technische Universität of Munich (TUM) for the GEM detectors [10], the Silicon detectors [11] and the new pixel GEM detectors [12]. A common project with the CEA Saclay has also permitted to develop a fast APV read-out for the multi-wire proportional chambers of the RICH detector [13]. However the front-end electronic cards have to be adapted to the Micromegas detectors, in particular concerning the protection and decoupling circuit in front of the APV.

An important feature of this electronics is its high density: an APV card reads 128 channels, and an ADC board, connected to 16 APV cards, reads 2048 channels. Only 17 APV cards and 2 ADC cards are needed. Moreover the cost of such electronics is rather low, less than 5 €/channel.

### 2.4 Improving robustness with bulk technology

Some of the old Micromegas detectors exhibited a few mechanical issues (gluing defaults, tightening of copper mesh, impurities below the mesh) concerning the board and the thin copper meshes, the board being built from thin 100 μm epoxy layers glued on a 5 mm thick honeycomb with the 5 μm copper mesh mounted on a frame.

The bulk technology is a way to improve the robustness of the detectors. A woven stainless steel mesh is laminated to the board between two photosensitive coverlays, and an UV insulation is applied with appropriate masks to draw pillars on the coverlays. At the end of the process the mesh is completely fixed to the board and is dust tight. It is also less sensitive to the gluing default of the epoxy layer to the honeycomb. The production of large size bulk Micromegas on honeycomb board has been studied in collaboration with the CERN TS-DEM laboratory and seems to be feasible.

A few aspects still need to be studied. The performances of detectors using thicker meshes with an hadron beam need to be carefully studied. In particular the discharge rate may be degraded by the higher thickness of the mesh. However a detector using similar meshes as in the bulk was used in COMPASS in 2006, and did not show more discharges. On the other hand, one also need to verify the compatibility of the bulk technology with honeycomb boards coated with a resistive layer.

## 3. Status of the project

### 3.1 High flux tests with present detectors

The pixelized Micromegas project has been launched in mid 2008. While first prototypes were being built, tests were realized in the COMPASS set-up using the old detectors. The goal was to control the gain stability of the detectors at the highest particle flux and the discharge rate with the high flux hadron beam. Two detectors were laterally shifted by 60 mm in order to have the full beam in their active area.

The COMPASS Micromegas detectors were tuned for the hadron beam used in 2008, with a modified gas mixture of Ne-$C_2H_6$-$CF_4$ 95-10-5%, and a mesh voltage reduced to 390 V in order to keep the discharge rate at an acceptable level (voltage set to 400 V with muon beam). With a muon beam of $10^7$ particles/s on a 40 cm long liquid hydrogen target the maximum counting rate was up to 230 kHz/channel, while this maximum rate went up to 130 kHz/channel for $4.10^6$ particles/s hadron beam. The discharge rate was between 0.2 to 0.3 discharges/s with the hadron beam and 390 V on the amplification gap, there was no discharge with the muon beam. In the future the nominal beam flux is expected to be about three times higher than during these tests.

The detector efficiency was measured in these conditions, giving values below 90% for channels which were seeing the beam (Figure 4). However the signal amplitudes, which are determined using the time over threshold measurements between the leading edge and the trailing edge of the signal, stayed the same as in normal conditions showing that there was no gain drop at the highest beam flux. The efficiency loss is fully due to the electronics occupancy.

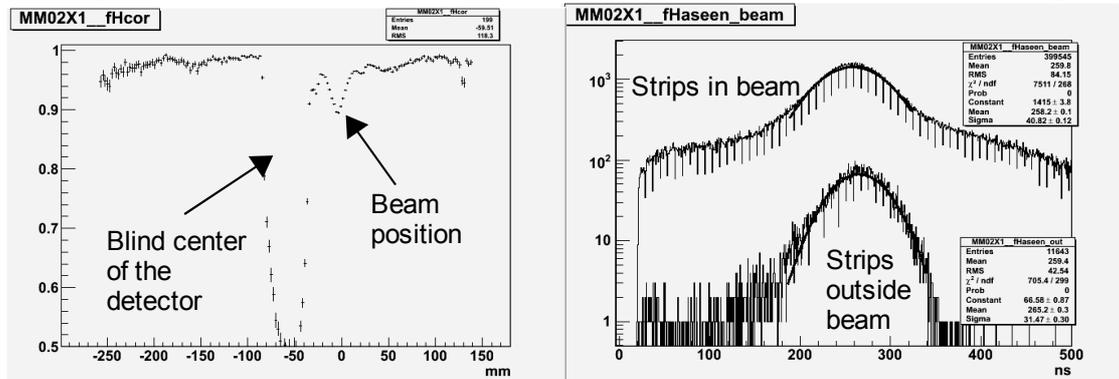

*Figure 4: Efficiency spectrum of an old Micromegas detector shifted to have the active area in the beam (left), the efficiency drop at -55 mm is due to the blind center of the detector, the inefficiency at the beam position reaches 10%. Time over threshold spectra of MM signals for strips in and outside the beam area (right). No gain drop is visible in the beam area, the tail at large time over threshold for strips in beam is due to pile-up effects.*

### 3.2 Prototypes of pixelized Micromegas

In order to test the concept of pixelized Micromegas, prototypes were built starting from an original design from the TUM Munich group adapted by the CERN TS-DEM laboratory and the CEA Saclay. They feature 32x32 1 mm² pixels in the center surrounded by 15 and 30 cm long strips (Figure 5 left). They are read by new APV cards with an adapted protection circuit.

Two prototypes were built, one with a 100 µm amplification gap and a 5 µm copper mesh mounted with the usual technology, the other one with the new bulk technology using a 30 µm stainless steel mesh (woven stainless steel mesh with 15 µm thread) and 128 µm amplification gap. The goal is to estimate the eventual impact of the bulk technology on the detector performances, in particular the discharge rate, the gain and also the spatial resolution.

The standard copper mesh prototype presents gain values similar to the old detectors, with no difference between pixels and strips. This detector was installed in the COMPASS set-up in front of the other Micromegas detectors and tested with hadron beams. With a similar flux and with 390 V on the amplification gap, the discharge rate was around 0.2 to 0.3 discharges/s, as expected based on the results from the shifted detector tests in 2008. New APV cards were delivered recently to equip all the strips and half of the pixels, efficiency and resolution measurements are not yet available.

The production of large size bulk Micromegas had never been done on honeycomb boards before. Several trials had to be done by the CERN TS-DEM laboratory before achieving a successful operation. The bulk prototype was finally delivered by the end of February, but several issues introduced delays: bad gluing, impurities introducing current leak and discharges.

This detector is working since June, and it was installed in the COMPASS set-up at the beginning of July. The first tests show that the discharge rate with hadron beams is similar to the copper mesh prototype for the same gain and beam flux. Efficiency and resolution measurements will be available soon.

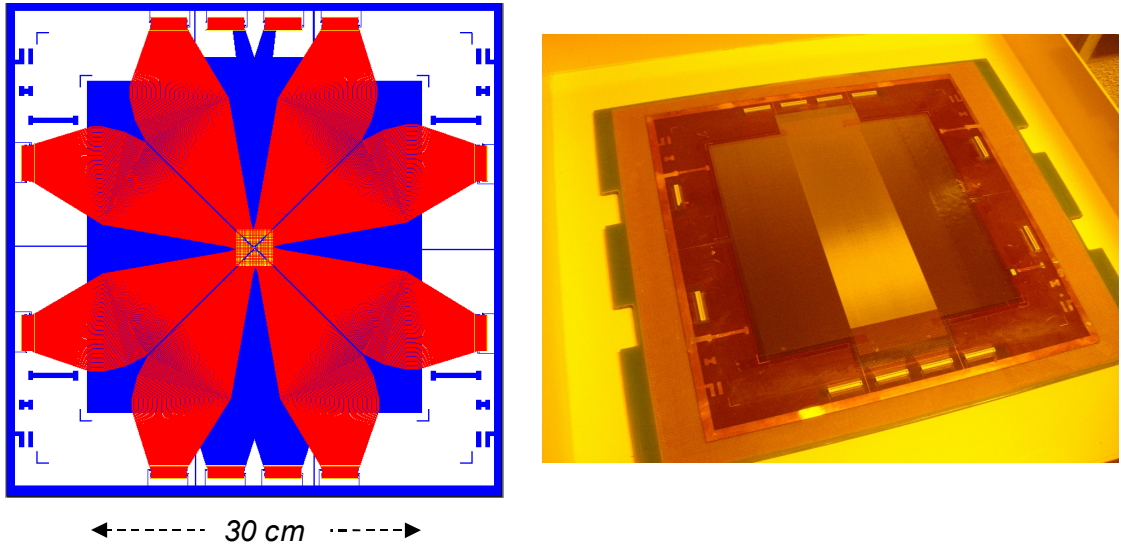

◄------- 30 cm -------►

*Figure 5: Scheme of the pixelized detector prototypes, with 32x32 1 mm² pixels in the center surrounded by 15 and 30 cm long strips (left). They feature a 2-layer board with pixel read-out routed following the original TUM design (in red). Picture of the Saclay pixelized bulk prototype (right)*

## 4. Perspective and time line

While studies on the two pixelized prototypes are going on, R&D on the resistive layers has started. Small prototypes (10x10 cm²) with several types of resistive layers will be tested during the RD-51 beam period in October in CERN H4 hadron beam line, in collaboration with other Saclay groups (in particular JLAB CLAS12, sLHC and ILC).

New 40x40 cm² prototypes featuring a pixelized area of 5 cm diameter with rectangular pixels will be designed and produced in the next months, to be tested in the COMPASS environment in 2010. R&D on resistive layers will be continued in parallel. The objective is to have a final prototype fulfilling all the proposed requirements in 2011, with the ultimate goal to equip the COMPASS spectrometer for the 2012 run.

## Acknowledgments

We would like to thank the E18 group of the TUM Munich university, and in particular Stephan Paul, for their great help. We also thank the CERN TS-DEM laboratory, especially Rui de Oliveira, Antonio Texeira and Olivier Pizzirusso. We acknowledge the important help of Alan Honma and Ian McGill of the CERN bonding laboratory for the production of the electronic readout cards.

# References

[1] COMPASS collaboration, *The COMPASS experiment at CERN*, Nucl. Instr. and Meth. A 577 (2007) 455

[2] D. Thers et al., *Micromegas as a large microstrip detector for the COMPASS experiment*, Nucl. Instr. and Meth. A 469 (2001) 133

[3] F. Kunne et al., *Micromegas: Large size high-rate trackers in the high energy experiment COMPASS*, proceedings of the NSS-IEEE 2006 conference, Nuclear Science Symposium Conference Record 2006-IEEE vol 6 (2006) 3838

[4] C. Bernet et al., *The 40cm x 40cm gaseous microstrip detector Micromegas for the high-luminosity COMPASS experiment at CERN*, Nucl. Instr. And Meth. A 536 (2005) 65

[5] E. Delagnes et al., *SFE16, a low noise front-end integrated circuit dedicated to the readout of large Micromegas detectors*, proceedings of the NSS-IEEE 1999 conference, IEEE Trans. Nucl. Sci. 47 (2000) 1447

[6] COMPASS collaboration, *Addendum 2 to the COMPASS Proposal*, CERN-SPSC-2009-025, SPSC-M-769, SPSLC-P-297 Add. 2, 21 June 2009

[7] COMPASS collaboration, *COMPASS Medium and Long Term Plans*, CERN-SPSC-2009-003, SPSC-I-238, 21 January 2009

[8] M.J. French et al., *Design and results from the APV25, a deep sub-micron CMOS front-end chip for the CMS tracker*, Nucl. Instr. and Meth. A 466 (2001) 359

[9] I. Giomataris et al., *Micromegas in a bulk*, Nucl. Instr. and Meth. A 560 (2006) 405

[10] B. Ketzer et al., *Performance of triple GEM tracking detectors in the COMPASS experiment*, Nucl. Instr. and Meth. A 535 (2004) 314

[11] H. Angerer et al., *Present status of silicon detectors in COMPASS*, Nucl. Instr. and Meth. A 512 (2003) 229

[12] T. Nagel et al., *A triple-GEM detector with pixel readout for high-rate beam tracking in COMPASS*, proceedings of the ICATPP 2007 conference, Villa Olmo, Como 8-12 October 2007

[13] P. Abbon et al., *A new analogue sampling readout system for the COMPASS RICH-1 detector*, Nucl. Instr. And Meth. A 595 (2008) 23